\documentclass[twocolumn]{aastex62}

\graphicspath{{./}{figures/}}

\usepackage{amssymb}
\usepackage{amsmath}

\shorttitle{Separating FRB using graph theory }
\shortauthors{Garc\'ia et al.}

\begin{document}

\title{Separating repeating fast radio bursts using the minimum spanning tree as an unsupervised methodology}

\correspondingauthor{C. R. Garc\'{i}a}
\email{crodriguez@ice.csic.es}

\author{C. R. Garc\'{i}a}
\affiliation{Institute of Space Sciences (ICE, CSIC), Campus UAB, Carrer de Can Magrans s/n, 08193 Barcelona, Spain}
\affiliation{Institut d’Estudis Espacials de Catalunya (IEEC), 08034 Barcelona, Spain}

\author{Diego F. Torres}
\affiliation{Institute of Space Sciences (ICE, CSIC), Campus UAB, Carrer de Can Magrans s/n, 08193 Barcelona, Spain}
\affiliation{Institució Catalana de Recerca i Estudis Avançats (ICREA), E-08010 Barcelona, Spain }
\affiliation{Institut d’Estudis Espacials de Catalunya (IEEC), 08034 Barcelona, Spain}

\author{Jia-Ming Zhu-Ge}
\affiliation{Nevada Center for Astrophysics, University of Nevada, Las Vegas, NV 89154, USA}
\affiliation{Department of Physics and Astronomy, University of Nevada, Las Vegas, NV 89154, USA}

\author{Bing Zhang}
\affiliation{Nevada Center for Astrophysics, University of Nevada, Las Vegas, NV 89154, USA}
\affiliation{Department of Physics and Astronomy, University of Nevada, Las Vegas, NV 89154, USA}



\begin{abstract}
Fast radio bursts (FRBs) represent one of the most intriguing phenomena in modern astrophysics. However, their classification into repeaters and non-repeaters is challenging. Here, we present the application of the graph theory Minimum Spanning Tree (MST) methodology as an unsupervised classifier of repeaters and non-repeaters FRBs.
By constructing MSTs based on various combinations of variables, we identify those that lead to MSTs that exhibit a localized high density of repeaters at each side of the node with the largest betweenness centrality.
Comparing the separation power of this methodology against known machine learning methods, and with the random expectation results, we assess the efficiency of the MST-based approach to unravel the physical implications behind the graph pattern.
We finally propose a list of potential repeater candidates derived from the analysis using the MST.
\end{abstract}


%\end{document}
\section{Introduction} \label{sec:intro}

Fast Radio Bursts (FRBs), first detected in 2001 by the Parkes Telescope and reported in 2007 \citep{Lorimer2007},
are transient radio pulses, typically in millisecond timescales and originating from cosmological distances.
FRBs have attracted considerable interest and numerous physical interpretations; see, e.g., \cite{Katz2018, Popov2018, Cordes2019, Petroff2019, Platts2019, Zhang2020, Xiao2021, Petroff2022, Xiao2022, Zhang2023}. 
The observed properties of FRBs, such as their dispersion (DM) and rotation (RM) measure, provide clues into the physical characteristics of the intervening medium, including density and magnetic field strength. 
Their energy or brightness temperature also offers valuable hints regarding the underlying mechanisms driving the FRB emission. 
It has been proposed that their emission could be powered by the dissipation of magnetic fields of a magnetar \citep{Popov2013, Lyubarsky2014, Katz2016, Lu2016, Murase2016, Kumar2017, Nicholl2017, Margalit2018, Lu2020, Yang2021}.
This has been substantiated after the Canadian Hydrogen Intensity Mapping Experiment (CHIME) \citep{CHIME} and STARE-2 \citep{STARE2} discovered FRB~200428 in association with a hard X-ray burst from the Galactic magnetar, SGR 1935+2154 \citep{HXMT-FRB, Mereghetti2020}.

After the discovery of the first repeater in 2016, FRB 20121102A \citep{Spitler2016, Scholz2016}, FRBs are broadly categorized into two groups based on their repetition properties: repeaters and (apparently) non-repeaters. 
In the magnetar interpretation, such differences may arise from age, with young magnetars producing a higher repetition rate than older ones, see e.g., \cite{Beloborodov2017, Metzger2017, Feng2022}.

However, synthesizing the physical causes behind these differences and escaping from often not fully understood observational biases remains a significant challenge.

Here, we aim to explore a novel classification tool for FRBs, separating them into repeaters and non-repeaters (for other approaches, see, e.g., \cite{Palaniswamy2018, Caleb2019, Ai2021, ML-FRBI, ML-FRBII}). 
We shall use graph theory (e.g., \cite{Wilson2010}) and principal component analysis (PCA, e.g., \cite{Pearson1901, Shlens2014}).
Graph theory aims to establish relationships between objects based on their connections and represent them in a graph.
In particular, we introduce a graph called the minimum spanning tree (MST, see recent astrophysical applications in \cite{MST-1, MST-2, Vohl2023}) as an unsupervised learning approach with a supervised evaluation.
The MST, where each node will represent an FRB, provides a manageable and compact structure calculated in an $N$-dimensional space.
Aided by the fundamental properties on which the MST is built, we will explore its capacity to establish the variables that show the best separation power and then indicate a group of likely repeating FRBs that have not yet been classified as such.
Using an unsupervised learning approach is particularly appealing when one does not control which properties within a data set are driving a classification, as is the case here. 
In addition, even the labels themselves are subject to possible change in time, as non-repeaters could become repeaters. 
Also, such methods do not necessarily require sample splitting for proper application.
This is valuable since the sample is not large and contains at most 750 FRBs, so we are not in the realm of big data. Randomly cutting the sample, in a training set (say, 80\%/20\% in size), would imply that 150 FRBs are sufficient to test the pattern behind a repeater/non-repeater classification, which we consider a risk.
Finally, the MST as an unsupervised method offers transparency in indicating the importance of the variables that best separate repeaters from non-repeaters. This becomes less clear in supervised methods when techniques such as feature importance (see eg., \cite{Feature_importance_1, Feature_importance_2, Feature_importance_3, Feature_importance_4}) or Shapley Additive exPlanations (SHAP, see \cite{SHAP_values1, SHAP_values2} for more details) are applied for that purpose.

%%%%%%%%%%%%%%%%%%%%%%%%%%%%%%%%%%%%%%%%%%%%%%%%%%
\section{Sample, variables, PCA, and MST}
\label{sec: sample_variables_PCA_MST}
%%%%%%%%%%%%%%%%%%%%%%%%%%%%%%%%%%%%%%%%%%%%%%%%%%

%%%%%%%%%% Sample
We use the sample from the Canadian Hydrogen Intensity Mapping Experiment Fast Radio Burst (CHIME/FRB) catalog \citep{CHIME_2021, CHIME_2023}.
We consider their 750 FRBs, with 265 as repeaters and 485 as non-repeaters. 
The group of repeaters corresponds to events coming from 70 localizations, and each FRB is taken individually as different events labeled as a repeater (as done in the CHIME catalog). 
Following the CHIME catalog, sub-bursts can be ascribed to both repeaters and non-repeaters, and they are also considered individual bursts, each with the corresponding label.
We exclude those six FRBs that have neither flux nor fluence measured.
%

%%%%%%%%%% Variables
To characterize an FRB we shall consider the logarithm of the following variables, 
(this is done just for convenience, as their values may differ by several orders of magnitude for different FRBs),
including the ones that are directly measured
\citep{CHIME_2021, CHIME_2023} such as peak frequency $\nu_{c}$ (MHz), flux $S_{\nu}$ (Jy), fluence $F_{\nu}$ (Jy ms), boxcar width $\Delta t_{BC}$ (ms), and also derived parameters based on the DM-$z$ relation (as presented in \citealt{Macquart_et_al2020, Deng2014, ML-FRBI, ML-FRBII}) such as redshift $z$, rest-frame frequency width $\Delta \nu$ (MHz), rest-frame width $\Delta t_{r}$ (ms), burst energy $E$ (erg), luminosity $L$ (erg/s), and brightness temperature $T_B$ (K).
As their distribution is not normal (nor is it that of the original variables without logarithm), we use the robust scaler to scale them (i.e., we subtract the median and divide by the interquartile range, see, e.g., \cite{NormTechniques})

\begin{eqnarray}
x_{i}^\dag = \frac{x_{i}-Q_{2}}{IQR}
~,\label{eq:robust_scaler}
\end{eqnarray}

In Eq. (\ref{eq:robust_scaler}) the $\dag$-symbol represents that the quantity $x_{i}$ has been scaled, $Q_{1}$, $Q_{2}$ and $Q_{3}$ represent the 1st quartile, median, and 3rd quartile of the distribution, respectively, and IQR is the interquartile range, $(Q_{3}-Q_{1})$.
%

%%%%%%%%%%%%%%%%%%%%%%%%%%%%%%%%%%%%%%%%%%%%%%%%%%%%%%%%%%%%%%%%%
%\section{Physical properties} 
%%%%%%%%%%%%%%%%%%%%%%%%%%%%%%%%%%%%%%%%%%%%%%%%%%%%%%%%%%%%%%%%%

\begin{figure*}
  \includegraphics[width=1\textwidth]{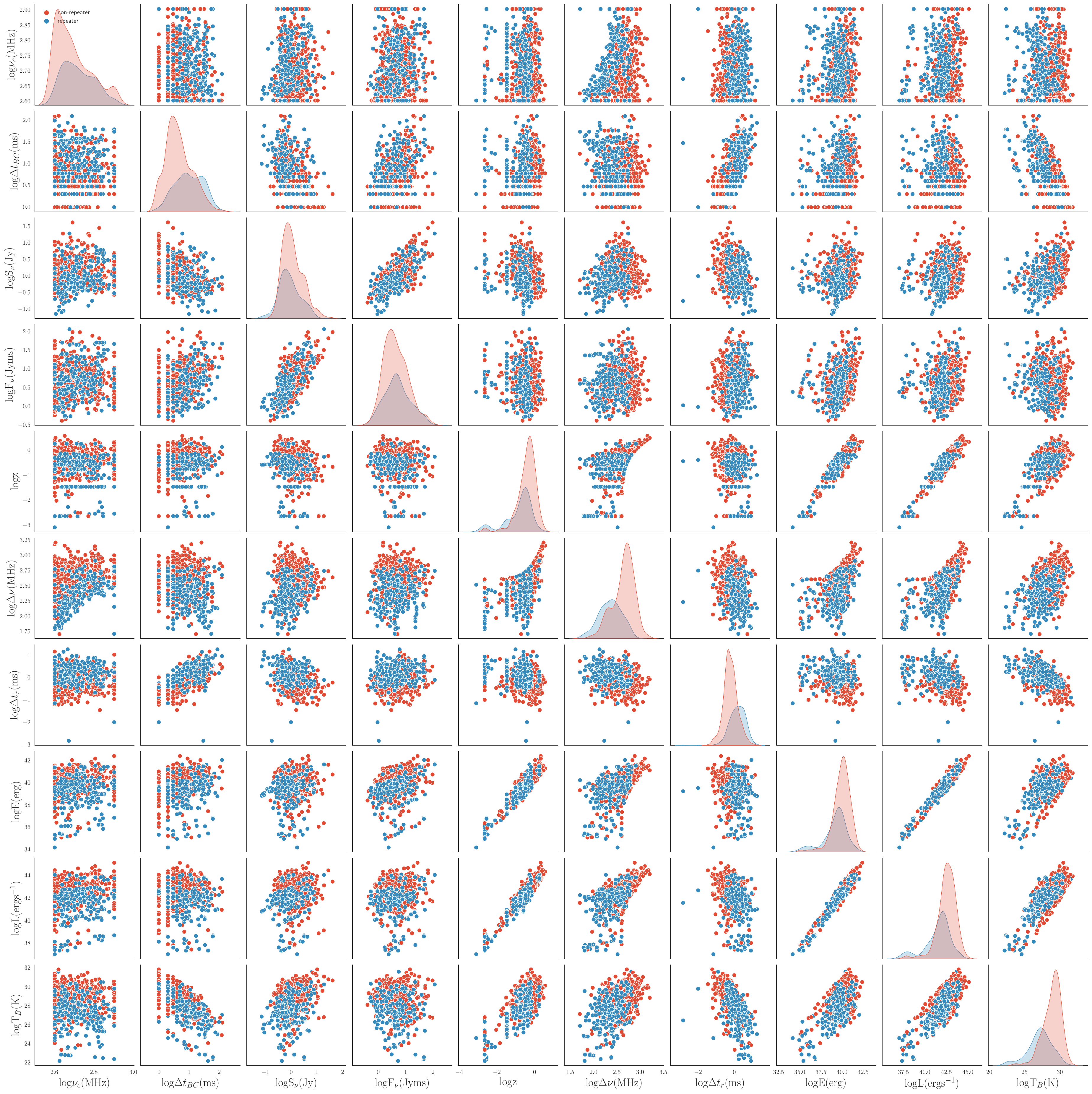}
  \centering
\caption{
Cross-correlation of the logarithm of the 10 magnitudes considered.
Repeaters (in blue) and non-repeaters (in red) are separately noted.
The main diagonal shows the distribution for each variable.
}
  \label{fig: pairplot_classes_originalVariables}
\end{figure*}

As we can observe with the cross-correlations shown in Fig. \ref{fig: pairplot_classes_originalVariables}, the physical properties have a different classification power.

This can be corroborated by applying a simple separation method: 
for each of the magnitudes considered, we note the median of each distribution and 
count the number of repeaters positioned on each side of this central location.
The best separation is provided by the rest-frame frequency width, which yields to
217 (82\%) grouped FRB repeaters (all with a width lower than 397.24 MHz).
This is far off from the random expectation: After producing 10$^5$ simulations in which repeater labels are randomly assigned, the average separation achieved is 52\%, that is, a separation consistent with a fully random process in which labels are separated in equal numbers on both sides of the median.
This high separation power of the rest-frame frequency width will 
serve as a threshold to validate the performance of other algorithms or classifiers.
We shall look below for classifiers that will perform better than 82\% in separating repeaters from non-repeaters.

\begin{figure*}
  \centering
  \includegraphics[width=1\textwidth]{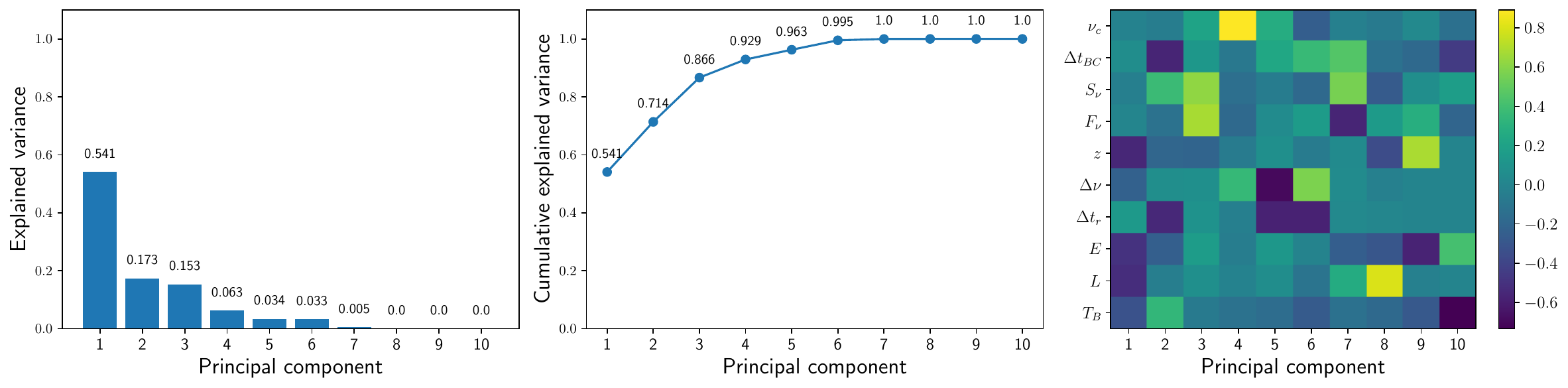}
  \caption{PCA results for the logarithm of the set of variables defined. 
  Left panel: The scree plot shows the explained variance of each PC. 
  Central panel: The cumulative explained variance of the PCs. 
  Right panel: The loadings of each variable regarding the PCs. 
  See the main text for further explanation.
}
  
  \label{fig: PCA}
\end{figure*}

%%%%%%%%%% PCA
Figure \ref{fig: PCA} presents the outcome of conducting PCA over the logarithm of all the variables defined in this section.
The left panel of Fig. \ref{fig: PCA}, also called the scree plot, shows the explained variance of each PC according to the eigenvalues of the covariance matrix.
Note that the covariance matrix calculates the relationships between pairs of variables, showing how changes in one variable are associated with changes in another.  
It represents the amount of information contained in each PC.
The central panel shows the cumulative explained variance of the PCs indicating how much of the total variance is captured as more PCs are included. 
The right panel shows the `weight', also called loading, that each variable has concerning each PC, indicating its contribution to the variance captured by that PC.
These values correspond to the coefficients of the eigenvectors of the covariance matrix. 
Negative values imply that the variable and the PC are negatively correlated. 
In contrast, a positive value shows a positive correlation between the PC and the variable.
While the first six PCs contain the majority of variance, retaining over 99\% of its informational content, the first three PCs contain 87\% of it.
Full variance coverage needs nine PCs, so the dimensionality reduction is not extreme if it is to be entirely retained. 
Thus, both the number of PCs needed to cover the total variance and the relative flat distribution of the loadings of the first PCs render its use a priori unappealing.

%%%%%%%%%% MST
Expanding and formalizing the work by \cite{Kruskal1956}, \cite{Kleinberg2005, Erickson2019} provide the foundational knowledge required to calculate the MST.
Considering the MST as a graph that connects points in a $N$-dimensional, each point (or node) is connected to at least one other end by an edge whose length is associated with a given distance. The edges of an MST are chosen so that the sum of their lengths is minimal and all nodes are connected. 
Our previous studies \cite{MST-1, MST-2}, also \cite{Vohl2023}, serve as a close example, as they discuss how these concepts are applied to a pulsar dataset.
Here, we shall introduce the Euclidean distance ($d$), which is calculated considering the straight-line distance between two nodes ($n, m$) in a $N$-dimensional space following the next expression
\begin{eqnarray}
d_{nm}=\sqrt{ \sum_{j=1}^{N} (n_{j}-m_{j})^{2} }
~.\label{eq: d_eucl.}
\end{eqnarray}
The $N$-distance seen in Eq. (\ref{eq: d_eucl.}), where $N$ represents the set of variables considered, can be computed using all or part of the defined physical variables mentioned above or the PCs (employing the PCs that account for 100\% of the explained variance yields an MST that is identical to the one derived using all magnitudes).
Taking the Eq. (\ref{eq: d_eucl.}), we initially generate a complete, undirected, weighted graph $G(V, E)$, characterized by a set $V$ of nodes (each of them is an FRB) with a size of $|V|$\footnote{The notation $|\cdot|$ represents the cardinality (or size) of the specified set, indicating the number of elements within it.} and set $E$ of edges of size $|E|$. The latter values are assigned to a specific weight $w$ equal to the Euclidean distance between the two FRBs, using all or part of the variables or PCs, as described.
For each combination of variables, then, we will start from the graph $G(750,280875)$ and derive an MST, $T(750,749)$. 
To do this, we apply the called Kruskal's algorithm (see eg., \cite{Kruskal1956}), which connects the whole sample considering the minimum distance between them at a local state, looking for a global minimization as the final condition (see the references above for a more detailed explanation and examples).

\section{Algorithm and results}

The total number of combinations (without repetition) from a set of 10 variables is 
$\sum_1^{10} \binom
{10}{k} = 2^{10} - 1 = 1023$,
reflecting the number of MSTs we compute.
For each of these MSTs, we identify the node with the highest centrality of the graph through the betweenness centrality estimator (see \cite{original, Moxley1974, Brandes, Baron}).
This estimator identifies how often a particular node appears on the shortest path between any two other nodes, effectively measuring the node's importance in terms of connectivity within the MST (see \cite{MST-2} and references therein for more details).
Eliminating this node would separate $T$ into several connected parts (or branches).
The branch with the highest number of repeaters will be named the repeater branch.
All other branches will be considered as non-repeater branches.
A high density of repeaters in the repeater branch would qualify the variables under which the underlying Euclidean distance is built as a well-separated filter via the MST algorithm.
To judge the performance of each MST seen as a classifier, we use the metrics: precision, recall, $F_{1}$, and $F_{2}$ score plus the area under the curve (AUC or ROC-AUC).
Precision measures the ratio of correctly identified repeaters regarding the total of FRBs (both repeaters and non-repeaters) of the repeater branch. At the same time, recall assesses the proportion of actual repeaters that are correctly identified. The $F_{1}$ score balances precision and recall equally, whereas the $F_{2}$ score places more emphasis on recall. AUC evaluates the overall discrimination ability of the classifier (see eg., \cite{han2011data} and the Appendix for more detailed definitions).
%
%%%%%% 
In identifying high-density locations of repeaters, metrics such as recall and precision focus on correctly identifying actual repeaters and minimizing non-repeaters taken as repeaters. 
A high $F_{1}$ score indicates a good balance between recall and precision. 
On the other hand, the $F_{2}$ score, as seen, places more emphasis on recall than precision, effectively tolerating a certain level of misclassification of non-repeaters as repeaters.
Lastly, a high AUC suggests that MST predictions rank true repeater instances higher than non-repeaters, which is crucial for identifying a high amount of repeaters among the whole sample.
To determine the best combination of variables, we look for a combination that offers a balanced trade-off between precision and recall (as reflected in the $F_{1}$ Score) and considers the $F_{2}$ Score (which weighs recall higher than precision) and the ROC-AUC value for overall performance.
This is in practice implemented by an overall rank (see, e.g., \cite{OverallRank2}), which averages all ranks across these evaluating metrics.

Armed with the MST/betweenness centrality methodology and with the rankings we find that there are 25 combinations of variables, and 3 combinations of PCs, for which we obtain a separating, i.e., recall, better than 82\%. 
Results are shown in Table \ref{table: final_results}.
Recall values are notably high across the board, which is excellent if the primary concern is to minimize missing repeaters. 
The precision is lower in some combinations than recall, which is expected in a scenario where the recall is prioritized.
The $F_{1}$ scores are still reasonably good, indicating a balanced performance between precision and recall. 
The $F_{2}$ scores are also high, reinforcing the model orientation towards recall which is desirable in contexts where the consequences of missing repeaters outweigh those of non-repeaters seen as repeaters.
The ROC-AUC values are also far from what could be considered a random guess, showing a good degree of separability.
Based on the overall rank, a combination of peak frequency, rest-frame frequency width, and brightness temperature (\#1) achieves the best balance.
Although this is not the best combination in terms of recall, for which it shows 0.8528, it is not far away from the best 0.8604 (a difference of just 2 repeaters) obtained with the combination (\#7).
Also, it shows a precision of 0.5825, being the second highest, which together with the aforementioned recall makes it the combination with the best $F_{1}$ score, $F_{2}$ score, and ROC value. 

The results obtained using the PCs are also shown in Table \ref{table: final_results}. 
Recalling that we need 9 PCs to define the full variance, we consider $\sum_1^{9} \binom{9}{k} =2^{9} - 1 = 511$.
Only three combinations of PCs present results exceeding the threshold taken for recall, although they are not particularly more informative or ranked better than those obtained via the direct use of physical variables.

We observe consistently high recall rates exceeding around 0.82 across all combinations, highlighting their effectiveness in accurately identifying repeaters.
The precision values, ranging approximately from 0.51 to 0.55, and corresponding $F_{1}$ scores between 0.63 to 0.67, suggest a reasonable balance between the accuracy of robust predictions and the method's ability to identify actual repeaters.
Furthermore, the $F_{2}$ scores, centered around 0.75, emphasize the evaluation's focus on recall over precision, aligning with our classification goal to reduce misclassified repeaters.
The ROC-AUC values, approximately 0.65 to 0.68, demonstrate the MST based on PCs' robust capability to distinguish between repeaters and non-repeaters, indicating significant discriminative power. 
Additionally, the high Z-scores of 10 and 12 across the combinations underscore the statistical significance of these results, affirming that the observed recall rates are not due to chance. 
Taking into account these points through the overall rank, the combination (PC1, PC2, PC3) emerges as the most effective. 
This represents 87\% of the total variance and has the best-separating power of all those using principal components. Having more variance described, with more PCs, does not yield a better separation power, implying that something else, other than the variables used to describe the sample here likely plays a role in the separation.

As before, we test the randomness of the process in two ways. 
First, we question whether, in case there is no link at all between repeating/non-repeating labels and the variables, we could still find a group of 25 combinations out of 1023 possible ones that could lead to a separating power over 82\%.
We do so by randomly relabelling the FRBs in repeaters and non-repeaters, maintaining the sample proportions, and building 10$^{5}$ fake samples for each MST using the same variables as the original.

We find that no combination of variables out of possible 1023 exceeds the threshold of 82\% separating in any of the simulations performed.
Second, for each combination selected in Table \ref{table: final_results} we compare the real result (i.e., its real separation power) with the average of the recall of the fake samples, using the Z-score, a statistical measure that quantifies how many standard deviations a data point (in this case, the recall) is from the mean of the distribution (see e.g., \cite{practicalStatistics} and the Appendix for more details).
Table \ref{table: final_results} shows that Z-scores obtained are at 9$\sigma$ or higher for recall values, along all combinations, compared to distributions from fake sets.
These randomness tests are a strong indicator of the effectiveness and robustness of this classification approach. 
In particular, the randomness of the best-ranked combination of variables, viewed in the Z-score of the first row in Table \ref{table: final_results}, is one of the lowest which proves a robust classification. 
%

%%%%%%%%%%%%%%%%%%%%%%%%%%%%%%%%%%%%% TABLES
\begin{table*}
\scriptsize
\centering
\caption{Summary of performance metrics of the combination of variables selected as those 
exceeding the threshold of 0.82 recall is seen in the rest-frame frequency width, sorted by overall rank.
The combinations seen in the 'variables' column are named as follows:
(1): $\mathrm{log}\nu_{c}$, 
(2): $\mathrm{log\Delta }t_{BC}$, 
(3): $\mathrm{log}S_{\nu}$, 
(4): $\mathrm{log}F_{\nu}$, 
(5): $\mathrm{log}z$, 
(6): $\mathrm{log}\Delta \nu$, 
(7): $\mathrm{log}\Delta t_{r}$, 
(8): $\mathrm{log}E$, 
(9): $\mathrm{log}L$, and 
(10):$\mathrm{log}T_B$.
}

\begin{tabular}{c|ccccccccccc}%{|l|l|l|l|l|l|l|l|l|l|l|l|}
\hline
\# & Variables & Precision & Recall & $F_{1}$ Score & $F_{2}$ Score & ROC-AUC & Z Score \\
\hline
\toprule
1 & [1, 6, 10] & 0.5825 & 0.8528 & 0.6922 & 0.7804 & 0.7773 & 14 \\
2 & [1, 5, 6, 7] & 0.5622 & 0.8528 & 0.6777 & 0.7729 & 0.7033 & 13 \\
3 & [1, 2, 5, 6, 10] & 0.5589 & 0.8415 & 0.6717 & 0.7642 & 0.7285 & 13 \\
4 & [1, 3, 4, 6, 7, 8, 9, 10] & 0.5784 & 0.8491 & 0.6881 & 0.7764 & 0.6834 & 13 \\
5 & [1, 5, 6, 8, 10] & 0.5878 & 0.8340 & 0.6895 & 0.7695 & 0.6961 & 14 \\
6 & [6] & 0.5722 & 0.8226 & 0.6749 & 0.7564 & 0.7522 & 19 \\
7 & [1, 2, 3, 6, 7, 8] & 0.5352 & 0.8604 & 0.6599 & 0.7672 & 0.6941 & 12 \\
8 & [1, 3, 5, 6, 9] & 0.5510 & 0.8566 & 0.6706 & 0.7711 & 0.6577 & 13 \\
9 & [1, 2, 3, 4, 6, 7, 9, 10] & 0.5457 & 0.8340 & 0.6597 & 0.7543 & 0.7007 & 12 \\
10 & [2, 5, 10] & 0.5598 & 0.8302 & 0.6687 & 0.7571 & 0.6999 & 13 \\%%%%
11 & [1, 3, 5, 6] & 0.5509 & 0.8377 & 0.6647 & 0.7587 & 0.6645 & 12 \\
12 & [1, 4, 5, 6, 9] & 0.5688 & 0.8264 & 0.6738 & 0.7578 & 0.6583 & 13 \\
13 & [5, 6, 9, 10] & 0.5198 & 0.8415 & 0.6427 & 0.7488 & 0.6883 & 11 \\
14 & [8, 10] & 0.5298 & 0.8377 & 0.6491 & 0.7505 & 0.6859 & 11 \\
15 & [2, 6, 8] & 0.5366 & 0.8302 & 0.6519 & 0.7483 & 0.7006 & 12 \\
16 & [1, 2, 4, 6, 7, 9, 10] & 0.5239 & 0.8264 & 0.6413 & 0.7409 & 0.7230 & 11 \\
17 & [1, 2, 4, 5, 6, 10] & 0.5201 & 0.8302 & 0.6395 & 0.7417 & 0.7118 & 11 \\
18 & [1, 4, 6, 9, 10] & 0.5116 & 0.8302 & 0.6331 & 0.7383 & 0.7013 & 10 \\
19 & [1, 2, 3, 6, 9, 10] & 0.5117 & 0.8226 & 0.6310 & 0.7335 & 0.7180 & 10 \\ %%%%
20 & [1, 2, 4, 5, 6, 9, 10] & 0.5034 & 0.8302 & 0.6268 & 0.7348 & 0.7002 & 10 \\
21 & [1, 2, 3, 4, 6, 7, 8, 9] & 0.4966 & 0.8377 & 0.6236 & 0.7366 & 0.6590 & 10 \\
22 & [1, 2, 4, 5, 6] & 0.4944 & 0.8340 & 0.6208 & 0.7332 & 0.6866 & 10 \\
23 & [1, 6, 8, 9, 10] & 0.4846 & 0.8302 & 0.6120 & 0.7266 & 0.7026 & 9 \\
24 & [1, 5, 6, 9, 10] & 0.4944 & 0.8302 & 0.6197 & 0.7309 & 0.6987 & 10 \\
25 & [1, 4, 5, 6, 9, 10] & 0.4943 & 0.8226 & 0.6176 & 0.7262 & 0.6851 & 10 \\
\hline
1 & [PC1,PC2,PC3] & 0.5504 & 0.8453 & 0.6667 & 0.7635 & 0.6858 & 12    \\ 
2 & [PC1,PC7] & 0.5178 & 0.8226 & 0.6356 & 0.7360 & 0.6716 & 11 \\ 
3 & [PC1,PC3,PC6,PC7,PC8] & 0.5092 & 0.8377 & 0.6334 & 0.7420 & 0.6523 & 11  \\ 
\hline
\end{tabular}
\label{table: final_results}
\end{table*}

\begin{figure*}
  \centering
  \includegraphics[width=1\textwidth]{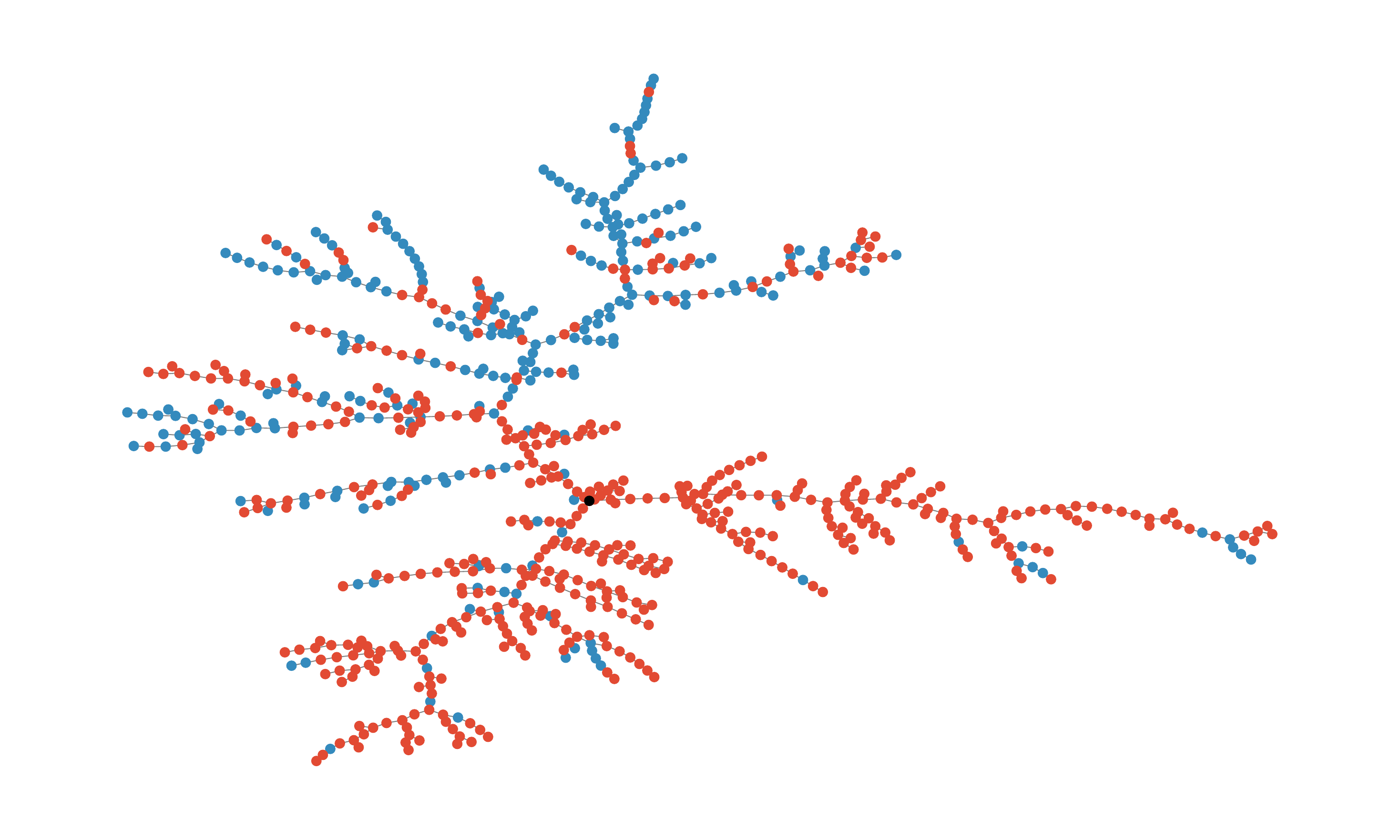}
  \caption{$T(750,749)$ computed from the Euclidean distance based on the combination of peak frequency, rest-frame frequency width, and brightness temperature. 
Repeaters are shown in blue and non-repeaters in red.
The most central node in terms of betweenness centrality appears in black.}
  \label{fig: MST_winner}
\end{figure*}

Combination \#1 successfully identified a significant number of repeaters (226/265), ensuring that relatively few repeaters are missed (39/265). 
Some non-repeaters are located in the repeater branch of the MST (162/485).
If there is any physical link between the variables and the repeating property, some of these should be candidates to appear as repeaters in the next catalog. We come back to this below.
Fig. \ref{fig: MST_winner} shows the MST with the arrangement of repeaters and non-repeaters of combination \#1, it separates the two types of FRBs.

\begin{table*}
\scriptsize
\centering
\caption{Properties of the currently labeled as non-repeater FRBs that are seen as repeaters for all combinations shown in Table \ref{table: final_results}.
The 'Sub' column shows the sub-burst number to which each FRB corresponds.
Candidate FRBs reported in \cite{ML-FRBI, ML-FRBII} are highlighted in black. 
Repeaters not classified as such in any of the selected combinations (outliers) are shown below the line.
}
\begin{tabular}{llcccccccccccc}
\hline
Name & Sub &
RA  & Dec   & $\nu_{c}$  & $\Delta t_{BC}$  & $S_{\nu}$  & $F_{\nu}$  & $z$ & $\Delta \nu$  & $\Delta t_{r}$  & $E$ & $L$  & $T_B$  \\
  &   &
 $(^\circ)$ &   $(^\circ)$ &  (MHz) &  (ms) & (Jy) &  (Jy ms) & &  (MHz) &  (ms) &  (erg) & (erg/s) &  (K) \\
\hline
\toprule
 FRB20180907E   &  0 & 167.88 & 47.09 & 400.20 & 11.80 & 0.73 & 6.90 & 0.3118 & 178.54 & 3.16 & $7.11        \times 10^{39}$ & $9.87 \times 10^{41}$ & $7.74 \times 10^{27}$ \\
 FRB20180920B   &  0 & 191.09 & 63.52 & 421.10 & 10.81 & 0.35 & 1.70 & 0.4007 & 116.53 & 1.66 & $3.10        \times 10^{39}$ & $8.93 \times 10^{41}$ & $6.71 \times 10^{27}$ \\
 FRB20180928A   &  0 & 312.95 & 30.85 & 400.20 & 2.95 & 1.34 & 2.50 & 0.0022 & 92.11 & 0.27 & $1.20          \times 10^{35}$ & $6.44 \times 10^{37}$ & $1.06 \times 10^{25}$ \\
\textbf{FRB20181017B}   &  0 & 237.76 & 78.50 & 593.20 & 12.78 & 1.06 & 6.50 & 0.2067 & 247.97 & 1.91 & $4.24\times 10^{39}$ & $8.35 \times 10^{41}$ & $1.86 \times 10^{27}$ 
 \\
 FRB20181022E  & 0 & 221.18  & 27.13 & 443.70 & 2.95 & 0.69 & 2.08 & 0.2073 & 193.28 & 0.33 & $1.02          \times 10^{39}$ & $4.09 \times 10^{41}$ & $4.09 \times 10^{28}$ \\
 FRB20181125A &  0 & 147.94 & 33.93 & 434.50 & 14.75 & 0.39 & 3.20 & 0.1710 & 156.33 & 1.09 & $1.04          \times 10^{39}$ & $1.48 \times 10^{41}$ & $6.49 \times 10^{26}$ \\
 FRB20181125A &  1 & 147.94 & 33.93 & 436.60 & 14.75 & 0.39 & 3.20 & 0.1710 & 177.76 & 1.23 & $1.04          \times 10^{39}$ & $1.49 \times 10^{41}$ & $6.43 \times 10^{26}$ \\
 FRB20181125A &  2 & 147.94 & 33.93 & 426.50 & 14.75 & 0.39 & 3.20 & 0.1710 & 141.34 & 1.35 & $1.02          \times 10^{39}$ & $1.45 \times 10^{41}$ & $6.73 \times 10^{26}$ \\
 FRB20181214A & 0 & 70.00   & 43.07 & 435.00 & 2.95 & 0.156 & 0.41 & 0.2308 & 116.19 & 0.43 & $2.47          \times 10^{38}$ & $1.15 \times 10^{41}$ & $1.20 \times 10^{28}$ \\
 FRB20181220A &  0 & 346.11 & 48.43 & 400.20 & 2.95 & 1.33 & 3.00 & 0.0022 & 196.64 & 0.43 & $1.44           \times 10^{35}$ & $6.39 \times 10^{37}$ & $1.05 \times 10^{25}$ \\
 FRB20181226E & 0 & 303.56  & 73.64 & 400.20 & 2.95 & 0.48 & 1.35 & 0.1779 & 186.23 & 0.99 & $4.37           \times 10^{38}$ & $1.83 \times 10^{41}$ & $2.55 \times 10^{28}$
 \\
 \textbf{FRB20181229B} &  0 & 238.37 & 19.78 & 445.50 & 20.64 & 0.42 & 4.90 & 0.3197 & 154.80 & 2.55 & $5.92 \times 10^{39}$ & $6.70 \times 10^{41}$ & $1.24 \times 10^{27}$ \\
 \textbf{FRB20190112A} & 0 & 257.98  & 61.20 & 697.70 & 9.83 & 1.40 & 16.20 & 0.3476 & 317.48 & 1.22 & $3.64 \times 10^{40}$ & $4.24 \times 10^{42}$ & $8.80 \times 10^{27}$ \\
 FRB20190128C &  0 &  69.80 & 78.94 & 491.60 & 15.73 & 0.71 & 5.90 & 0.1772 & 238.50 & 5.23 & $2.32          \times 10^{39}$ & $3.29 \times 10^{41}$ & $8.73 \times 10^{26}$ \\
 FRB20190206B & 0 & 49.76   & 79.50 & 506.40 & 19.66 & 0.95 & 9.60 & 0.2190 & 350.34 & 5.82 & $6.03          \times 10^{39}$ & $7.27 \times 10^{41}$ & $1.09 \times 10^{27}$ \\
 \textbf{FRB20190206A} &  0 & 244.85 &  9.36 & 534.50 & 5.90 & 1.40 & 9.10 & 0.0618 & 213.84 & 0.76 & $4.53  \times 10^{38}$ & $7.40 \times 10^{40}$ & $1.20 \times 10^{27}$ \\
 \textbf{FRB20190218B} &  0 & 268.70 & 17.93 & 588.00 & 17.69 & 0.57 & 5.90 & 0.4416 & 334.17 & 1.42 & $1.84 \times 10^{40}$ & $2.56 \times 10^{42}$ & $2.56 \times 10^{27}$ \\
 FRB20190223A & 0 & 64.72   & 87.65 & 444.80 & 3.93 & 0.47 & 1.58 & 0.2865 & 149.61 & 0.59 & $1.52           \times 10^{39}$ & $5.81 \times 10^{41}$ & $3.05 \times 10^{28}$ \\
 FRB20190308C &  0 & 188.36 & 44.39 & 453.40 & 21.63 & 0.47 & 4.80 & 0.4542 & 218.85 & 0.28 & $1.22          \times 10^{40}$ & $1.74 \times 10^{42}$ & $2.51 \times 10^{27}$ \\
 FRB20190308C &  1 & 188.36 & 44.39 & 449.00 & 21.63 & 0.47 & 4.80 & 0.4542 & 211.58 & 0.38 & $1.21          \times 10^{40}$ & $1.72 \times 10^{42}$ & $2.56 \times 10^{27}$ \\
 FRB20190323D &  0 &  56.88 & 46.93 & 400.20 & 12.78 & 0.37 & 2.49 & 0.5930 & 204.86 & 3.41 & $9.66          \times 10^{39}$ & $2.29 \times 10^{42}$ & $1.26 \times 10^{28}$ \\
 \textbf{FRB20190329A} &  0 &  65.54 & 73.63 & 432.30 & 11.80 & 0.52 & 2.24 & 0.0022 & 73.87 & 1.04 & $1.16  \times 10^{35}$ & $2.70 \times 10^{37}$ & $2.20 \times 10^{23}$ \\
 \textbf{FRB20190410A} &  0 & 263.47 & -2.37 & 515.70 & 6.88 & 1.59 & 5.80 & 0.0728 & 182.92 & 0.94 & $3.89  \times 10^{38}$ & $1.14 \times 10^{41}$ & $1.51 \times 10^{27}$ \\
 \textbf{FRB20190412B} &  0 & 285.65 & 19.25 & 400.20 & 42.27 & 0.68 & 12.80 & 0.0146 & 228.59 & 6.70 & $2.60\times 10^{37}$ & $1.40 \times 10^{39}$ & $1.11 \times 10^{24}$ \\
 \textbf{FRB20190423B} &  0 & 298.58 & 26.19 & 537.60 & 9.83 & 0.87 & 7.00 & 0.0031 & 159.79 & 2.48 & $8.54  \times 10^{35}$ & $1.06 \times 10^{38}$ & $6.48 \times 10^{23}$ \\
 \textbf{FRB20190423B} &  1 & 298.58 & 26.19 & 524.60 & 9.83 & 0.87 & 7.00 & 0.0031 & 148.96 & 8.47 & $8.33  \times 10^{35}$ & $1.04 \times 10^{38}$ & $6.81 \times 10^{23}$ \\
 \textbf{FRB20190429B} &  0 & 329.93 &  3.96 & 422.40 & 16.71 & 0.74 & 5.00 & 0.1944 & 50.64 & 5.34 & $2.05  \times 10^{39}$ & $3.62 \times 10^{41}$ & $1.32 \times 10^{27}$ \\
 FRB20190430A &  0 &  77.70 & 87.01 & 433.80 & 19.66 & 0.75 & 7.70 & 0.2278 & 214.13 & 2.75 & $4.50          \times 10^{38}$ & $5.38 \times 10^{41}$ & $1.27 \times 10^{27}$ \\
 \textbf{FRB20190527A} &  0 &  12.45 &  7.99 & 484.70 & 57.02 & 0.47 & 10.10 & 0.5367 & 205.46 & 1.74 & $3.87\times 10^{40}$ & $2.77 \times 10^{42}$ & $4.46 \times 10^{26}$ \\
 FRB20190527A &  1 &  12.45 &  7.99 & 449.10 & 57.02 & 0.47 & 10.10 & 0.5367 & 172.11 & 1.61 & $3.59         \times 10^{40}$ & $2.56 \times 10^{42}$ & $5.20 \times 10^{26}$ \\
 FRB20190601C &  0 &  88.52 & 28.47 & 517.00 & 5.90 & 1.32 & 5.80 & 0.1753 & 223.54 & 0.58 & $2.35           \times 10^{39}$ & $6.28 \times 10^{41}$ & $1.02 \times 10^{28}$ \\
 FRB20190601C &  1 &  88.52 & 28.47 & 502.20 & 5.90 & 1.32 & 5.80 & 0.1753 & 201.91 & 0.43 & $2.28           \times 10^{39}$ & $6.10 \times 10^{41}$ & $1.08 \times 10^{28}$ \\
 FRB20190617B &  0 &  56.43 &  1.16 & 459.30 & 13.76 & 0.99 & 9.20 & 0.1655 & 217.37 & 6.50 & $2.94          \times 10^{39}$ & $3.69 \times 10^{41}$ & $1.58 \times 10^{27}$ \\
 \hline
FRB20180910A &  0 & 354.83 & 89.01 & 417.60 & 0.98 & 6.50 & 5.60 & 0.6230 & 649.22 & 0.126 & $2.51 \times 10^{40}$ & $4.73 \times 10^{43}$ & $3.82 \times 10^{31}$ \\
FRB20190210C &  0 & 313.90 & 89.19 & 448.50 & 1.97 & 2.37 & 3.60 & 0.5798 & 631.93 & 0.181 & $1.50 \times 10^{40}$ & $1.56 \times 10^{43}$ & $2.58 \times 10^{30}$ \\
FRB20200726D &  0 & 294.75 & 59.40 & 684.00 & 4.92 & 0.76 & 3.48 & 1.3812 & 541.72 & 0.407 & $1.23 \times 10^{41}$ & $6.37 \times 10^{43}$ & $3.17 \times 10^{29}$ \\

\hline
\end{tabular}
\label{table: candidates}
\end{table*}

\subsection{Results considering selection effects}

As explained in \cite{CHIME_2021}, assessing selection effects is a challenge to be addressed in the study of FRBs. 
We have approached this question following the considerations in Section 3.3 of \cite{CHIME_2023}. The following cuts in the sample have been imposed

\begin{enumerate}
    \item Events measured by the \textit{bonsai} real-time detection pipeline S/N $<$ 12 are excluded due to being more likely misclassified as noise.
    \item Events with DM $<$ 1.5 max(DM$_{\mathrm{NE2001}}$, DM$_{\mathrm{YMW16}}$) are excluded to minimize the possibility of having a wrong identification of rotating radio transients or radio pulsars as FRBs considering that a $\sim 50\%$ errors may be common in models used to estimate the Galactic DM.
    \item Events detected in the telescope's sidelobes (see \cite{Lin_sideLobes1, Lin_sideLobes2}) are excluded since the understanding of the primary beam shape at large zenith angles is limited. This makes it difficult to characterize events. 
\end{enumerate}

We are left with 459 FRBs after considering such cuts. This is a significant reduction from the earlier sample (by about 40\%). In this sample, we find 135 FRBs classified as repeaters. 
For each of the variables considered, again, we note the median of each distribution and count the number of repeaters positioned on each side of this central location. The best classification is provided once again for this reduced sample by the rest-frame frequency width, which can separate 109 of the 135 repeaters (80\%).
Thus, we test whether our methodology provides a separating power that goes beyond this threshold.
After applying it similarly to what has been done for the full sample we find 15 different combinations of variables 
for which the separating performance exceeds the one provided by the rest-frame frequency width.

We note two aspects of interest, selection effects may indeed play a role both forcing to work with reduced samples whose properties are likely not well mapped yet due to low-number statistics, and changing the total variance. As a result, the PCA analysis and the combination of variables for which the separating power is high are different when compared with those corresponding to the full sample. 
However, it can be recognized that peak frequency, rest-frame frequency width, and brightness temperature (variables in \#1, see Table \ref{table: final_results}) are among the variables that appear the most in all combinations with large separating power, whereas, on the other hand, the burst energy is the least implied, offering little classification power. 

%%%%%%%%%%%%%%%%%%%%%%%%%% %%%%%%%%%%%%%%%%%%%%%%%%%% 
\section{Conclusions}
%%%%%%%%%%%%%%%%%%%%%%%%%% %%%%%%%%%%%%%%%%%

To our knowledge, this is the first application of the graph theory MST technique to separate the repetition properties of FRBs.
To some extent, this work relates to that of \cite{Bhatporia2023} when using mathematical techniques to classify the population of the FRBs, although in their case performance metrics are not provided.
Our performance metrics indicate that the MST classifiers do a good job of separating repeaters among the FRBs, excelling especially in recall and thus minimizing false negatives.
This approach demonstrates the potential for practical application and can contribute to the scientific understanding of FRBs by highlighting the importance of certain variables over others concerning what best describes the repetitive character of a source.

Methodologically, we are not defining clusters over the MST, we compute instead the betweenness centrality for each node and identify the one with the highest value (the black node in Fig. \ref{fig: MST_winner}). 
This node acts as a bridge within the MST, by removing it, we partition the graph into separate branches. We then use the labels to count the number of repeaters on each branch, which allows us to study the distribution of repeaters and non-repeaters. 
Since we do not cluster, we do not confront such issues as chaining or unbalanced clustering seen, for example, in single linkage algorithms (see \cite{Everit_Landau_clusters} for a discussion). 
Importantly, we do not use the labels during the construction of the MST or the identification of the most central node; our method is unsupervised and does not require any training. 
As a result, the MST works with the whole sample at once, without the need to divide the sample into smaller pieces and risk losing relevant information in the process.

Table \ref{table: final_results} shows that while rest-frame frequency width and brightness temperature demonstrated strong discriminatory power per se via median separation of the sample, the frequent occurrence of peak frequency in the selected combinations highlights its potential contribution to the classification process. 
This information can be valuable not only for optimizing the current method but also for informing feature selection strategies in other classification methods, for instance, using machine learning.

The performance of this method is on par or exceeds that of others. For instance, 
\cite{Chen_2021, ML-FRBI, ML-FRBII} use the sample from \cite{CHIME_2021} under a machine learning approach.
We apply our methodology in the same conditions of \cite{ML-FRBI, ML-FRBII}, where the sample is composed of 594 FRBs, where 500 non-repeaters and 94 repeaters are noted.  
Under these assumptions, the results are a 97\% recall and an $F_{2}$ score of 77\%, which is comparable to that obtained in the quoted papers.
The recall is higher than those seen in \cite{Chen_2021}, where the sample is the same but they identify one repeater less (501 non-repeaters and 93 repeaters).
In this work, the variable set adds more properties to deal with a total of 13 (10 observational and 3 model-dependent parameters, as they define), among which are several of those used in our set.

In Table \ref{table: candidates} we show 33 currently labeled as non-repeaters FRBs that appear in the repeater branch identified using the 25 selected combinations obtained of variables showing a separating power exceeding a recall threshold of 82\%.
In the context of the MST strategy, these are strong candidates for repeating bursts.
A few have been identified as well using machine learning methodologies in \cite{ML-FRBI, ML-FRBII} 
Table \ref{table: candidates} also shows the few FRBs denoted as outliers, those that are known repeaters but are consistently classified 
in the non-repeater branch of any of the 25 MSTs considered. 
Both candidates and outliers constitute a handle on the power and caveats of the technique, to keep an eye on with future samples of FRBs.

Reducing false positives can be important to gain robustness, and attempts could be made to improve precision without significantly compromising recall. 
To refine the classification, a methodology that involves dividing the graph into significant branches (as introduced in \cite{MST-2}) could be of help. 

Finally, one obvious enhancement we would like to perform provided the opportunity is to consider the rate of repetition of each of the repeating FRBs.
In particular, we would like to see whether the MST grouping can distinguish a soft transition across the sample from the most commonly repeating FRBs to the non-repeater ones. 
However, this exercise is hindered by the current availability of data because such a rate should consider the right integration time for a particular field of view and this information is not provided in the catalog information \cite{CHIME_2021, CHIME_2023}. 
We encourage such data to be provided in future versions of the catalog, as it would be beneficial for this as well as similar analysis, and of course for comparing with any modeling of the repeating mechanism.

%%%%%%%%%%%%%%%%%%%%%%%%%%%%%%%%%%%%%%%%%%%%%%%%%%
\section*{Acknowledgements}
%%%%%%%%%%%%%%%%%%%%%%%%%%%%%%%%%%%%%%%%%%%%%%%%%%

This work was supported by the grant PID2021-124581OB-I00 of MCIU/AEI/10.13039/501100011033 and 2021SGR00426. 
This work was also supported by the program Unidad de Excelencia María de Maeztu CEX2020-001058-M and by MCIU with funding from the European Union NextGeneration EU (PRTR-C17.I1). C. R. García thanks UNLV for its hospitality during part of this investigation.

\bibliographystyle{aasjournal}
\bibliography{biblio}

%%%%%%%%%%%%%%%%%%%%%%%%%%%%%%%%%%%%%%%%%%%%%%%%%%
\section*{Appendix
\label{sec:appendix}}
%%%%%%%%%%%%%%%%%%%%%%%%%% Appendix %%%%%%%%%%%%%%%%%

\textit{Precision} quantifies the ratio of repeaters that were correctly identified concerning the total in the branch, while 
{\it Recall} (also known as {\it sensitivity}) measures the proportion of actual repeaters that were correctly classified.
The F measure, defined as
\begin{equation}
F_{\beta} = (1 + \beta^2)  \frac{Precision \,\, Recall}{(\beta^2 \,\, Precision) + Recall}
\end{equation} evaluates the predictive performance by combining precision and recall through a weighted harmonic mean. 
The importance of precision ($\beta < 1$) versus recall ($\beta > 1$) is modulated by the $\beta$ parameter.
When $\beta = 1$, precision and recall are equally weighted. 
The $F_{1}$ score is obtained when $\beta = 1$, therefore leading to an equal emphasis on precision and recall.
The $F_{2}$ score, alternatively, is derived when $\beta = 2$, giving more importance to recall over precision. 
The \textit{Receiver Operating Characteristic (ROC)} curve is a graphical representation of the performance of a binary classification model across different threshold settings. 
It plots the true positive rate (sensitivity) against the false positive rate (1 - specificity) at various threshold values. 
Here, \textit{Specificity}, also known as the true negative rate, measures the proportion of actual negative instances (non-repeater FRBs) correctly identified as negative.
The ROC curve helps assess the trade-off between sensitivity and specificity and provides insights into the classifier's ability to discriminate between positive and negative instances.
The \textit{Area Under the Curve (AUC)} value quantifies the overall performance of a binary classification model represented by the ROC curve. 
It measures the area under the ROC curve, with values ranging from 0 to 1. 
AUC values closer to 1 indicate better discrimination ability of the classifier, while values closer to 0.5 suggest poor discrimination (similar to random guessing).

The labels assigned to repeaters and non-repeaters according to the repeater branch are as follows:
\begin{itemize}
  \item \textbf{True Positive (TP)}: Repeaters correctly classified as repeaters 
  \item \textbf{False Positive (FP)}: Non-repeaters incorrectly classified as repeaters.
  \item \textbf{True Negative (TN)}: Non-repeaters correctly classified as non-repeaters.
  \item \textbf{False Negative (FN)}: Repeaters incorrectly classified as non-repeaters.
\end{itemize}
with these, the scores are defined as
  \begin{eqnarray}
    Precision &=& \frac{TP}{TP + FP}\\
    Recall &=& \frac{TP}{TP + FN} \\
    Specificity &=& \frac{TN}{TN + FP}\\
    F_{1}\; Score &=& \frac{2 \,\, Precision \,\, Recall}{Precision + 
    Recall}\\
    F_{2}\; Score &=& \frac{5 \,\, Precision \,\, Recall}{4 \,\, Precision + 
    Recall}
  \end{eqnarray}
The Z-score, also known as the standard score or z-value, is a statistical measure that quantifies how many standard deviations a data point is from the mean of a dataset. It is defined as 
\begin{eqnarray}
    Z = \frac{x - \mu}{\sigma}.
\end{eqnarray}
Here $x$ is the value of the data point (the classifier's recall), $\mu$ and $\sigma$ are the mean and the standard deviation of the dataset (distribution from the fake cases), respectively. 
A Z-score of 0 indicates that the data point is exactly at the mean of the dataset, while positive and negative Z-scores indicate that the data point is above and below the mean, respectively.
A Z-score higher than 2$\sigma$ or 3$\sigma$ in absolute value is significant at the 5\% or even 1\% level.

\end{document}